\documentclass[conference]{IEEEtran}
\IEEEoverridecommandlockouts

\usepackage{cite}
\usepackage{amsmath,amssymb,amsfonts}
\usepackage{algorithmic}
\usepackage{graphicx}
\graphicspath{{figs/}}
\usepackage{textcomp}
\usepackage{xcolor}
\usepackage{booktabs}
\usepackage{array}
\usepackage{url}
\usepackage{hyperref}

\def\BibTeX{{\rm B\kern-.05em{\sc i\kern-.025em b}\kern-.08em
    T\kern-.1667em\lower.7ex\hbox{E}\kern-.125emX}}

\begin{document}

\title{Prices, Probabilities, and Parlays: Systematic Bias in Sports Prediction Markets}

\author{
  Niusha Moshrefi\\
  Princeton University, Princeton, NJ, USA\\
  \texttt{niusha@princeton.edu}
}
\maketitle

\begin{abstract}
Prediction market prices are routinely interpreted as probabilities, both in academic work and in derivative products built atop these markets. We document two systematic ways this interpretation fails on Kalshi, a major U.S. event-contract exchange, using 23 million moneyline trades across major sport leagues. First, calibration, the agreement between quoted prices and realized event frequencies, is not a static property of a contract: fitting calibration models within time-to-expiry buckets, we find that parameters sit at their perfect-calibration reference values in the middle of a contract's life but depart sharply as expiry approaches. In the final ten minutes before settlement the empirical calibration curve becomes step-like, fitting a Prelec form with curvature parameter well above one — the opposite sign of the canonical lottery-choice fit, consistent with insurance-demand behavior by traders holding losing positions. Second, cross-game parlays on Kalshi are systematically overpriced relative to the product of their contemporaneous leg prices, with overpricing growing in leg count. This holds when the parlay legs are drawn from the TTE regime in which leg-level calibration is essentially perfect, indicating a separate, market-level markup at the parlay-pricing stage. Both deviations are systematic and therefore admit computational correction. Practical use of prediction-market prices as probabilities requires conditioning on time to expiry and on product type, not on price alone.
\end{abstract}


\section{Introduction}\label{sec:intro}
Prediction markets have moved from academic curiosity to mainstream financial instrument. The two largest event-contract platforms, Kalshi and Polymarket, collectively passed \$150 billion in lifetime trading volume in April 2026~\cite{defillama2026}. A foundational claim underwrites this growth, that the market price of a binary contract paying \$1 on event A can be read directly as the market's probability estimate of A~\cite{wolfers2004prediction}. This interpretation is invoked by traders sizing positions, by researchers using prices as belief proxies, and increasingly by mainstream media — the \textit{Wall Street Journal}, through its data partnership with Polymarket, has run headlines of the form ``Polymarket Bets See Over 70\% Chance of U.S.\ \dots''~\cite{intercept2026} — and, most consequentially for this paper, by the construction of derivative products such as parlays whose payoffs are explicit functions of underlying contract prices.

This paper argues that the price-as-probability interpretation is systematically wrong in a way that matters for downstream applications, and it does so by examining a dimension of miscalibration that has received little attention: time to expiry. A prediction market is calibrated if contracts trading at price \(p\) resolve in favor of the event with frequency \(p\). Most of the prediction-market bias literature treats calibration as a static property of a market — a single calibration curve fit to all trades — and documents level effects such as the favorite-longshot bias \cite{snowberg2010explaining,ottaviani2008favorite}. We instead slice the data by time remaining until contract settlement and show that the calibration function itself is a non-trivial function of that time. The market is not simply biased; the structure of its bias evolves predictably as expiry approaches.

\begin{figure*}[t]
  \centering
  \includegraphics[width=0.9\textwidth]{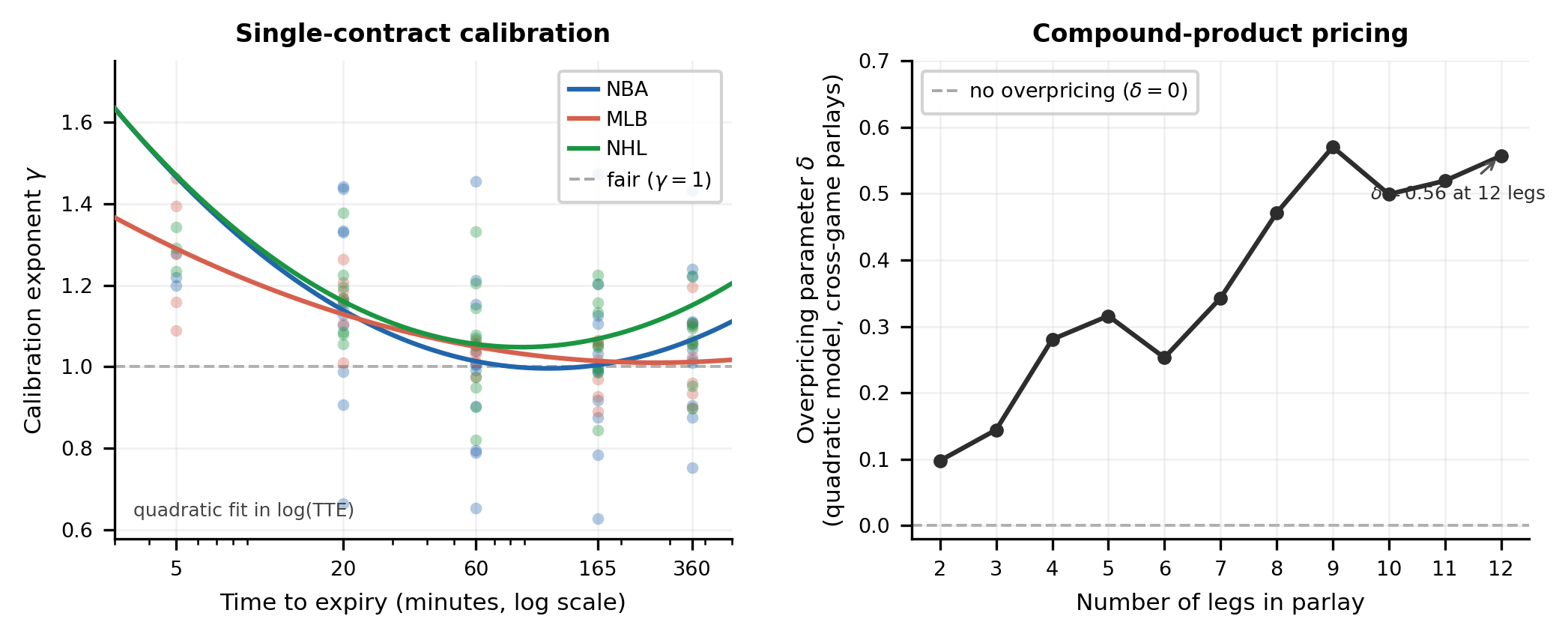}
  \caption{\textbf{Left:} Single-contract
calibration exponent \(\hat{\gamma}\) versus time-to-expiry, pooled
per league; \(\hat{\gamma} \approx 1\) in the middle of a contract's
life and rises sharply near settlement. \textbf{Right:} Overpricing
parameter \(\hat{\delta}\) from \(\hat{C}(p) = p - \delta\, p(1-p)\),
fit per leg count on cross-game parlays whose legs lie in the
calibrated regime on the left.}
  \label{fig:headline}
\end{figure*}

Figure~\ref{fig:headline} previews the two empirical findings around which the paper is organized: single-contract calibration depends sharply on time-to-expiry (left panel), and the parlay product built on top of these contracts is systematically overpriced in a leg-count-dependent way (right panel).

We make three contributions. 
\begin{itemize}
    \item First, we document that calibration curves in sports prediction markets vary systematically with time-to-expiry, and we characterize this variation parametrically. Fitting a quadratic-in-logit specification to trades grouped by time-to-expiry bucket, we find that the curve parameters themselves trace out a U-shape. Contracts traded well before tip-off or first pitch, and contracts traded near the final whistle, are the most miscalibrated, while trades during the bulk of in-game play lie closest to the diagonal. This pattern is robust across three major leagues — the NBA, NHL, and MLB — suggesting a structural feature of how these markets aggregate information rather than a quirk of any one sport.
    \item Second, we show that calibration curves near expiry are not merely more biased but qualitatively different in shape: they become Prelec-like \cite{prelec1998}, consistent with a probability-weighting distortion of the kind documented in cumulative prospect theory \cite{tversky1992advances}. We interpret this as evidence of insurance-demand behavior — traders holding losing positions late in a game face increasingly convex losses and hedge by paying premiums on the opposing side — and we discuss alternative explanations that our data cannot fully rule out.
    \item Third, we show that even when single-leg prices are well-calibrated, derivative products built from them are not. Treating cross-game parlays as a market-implied estimate of a joint probability, we find that parlay prices on Kalshi are systematically too high relative to the product of contemporaneous leg prices, with median overpricing growing geometrically in the number of legs. We show parlay overpricing cannot be inherited from leg-level miscalibration; together with the cross-game restriction that substantially reduces dependence, it points to a separate, market-level markup imposed at the parlay-pricing stage.
\end{itemize}

The implications run in two directions. For practitioners and researchers using prediction-market prices as probability inputs, our results indicate that any such use should condition on both time-to-expiry and product type — a single-leg price in the middle of a contract's life behaves like a probability, while the same contract in its final minutes or assembled into a parlay does not. For market microstructure, the time-evolution of the calibration function offers a new empirical handle on the behavioral and informational forces shaping price formation in event-contingent contracts.


\section{Related Work}\label{sec:related-work}

Our paper sits at the intersection of two literatures: information aggregation in prediction markets, and documented systematic biases in prediction and sports-betting prices. We discuss each in turn and position our contribution.

\paragraph{Information aggregation in prediction markets} A foundational line of work argues that markets in event-contingent contracts efficiently aggregate dispersed private information, yielding prices that can be interpreted as market-aggregated probability forecasts~\cite{wolfers2004prediction,wolfers2006five}. Empirical studies have generally found that such forecasts are accurate on average and outperform standard polling or expert benchmarks across a range of domains, from political elections to corporate sales forecasts~\cite{wolfers2004prediction}. This literature supplies both the practical motivation for the price-as-probability interpretation we examine and the empirical baseline against which we measure deviations. Closer to our downstream analysis, a smaller but growing literature studies how prediction-market mechanisms can be extended to joint events. Most recently, Rana et al.~\cite{rana2026parlaymarket} propose an automated market-making design that prices parlay-style joint contracts within a unified liquidity pool, representing the joint distribution via a pairwise exponential-family model. Our work is complementary: rather than designing a coherent mechanism for joint pricing, we measure how badly existing platforms misprice joint contracts when leg prices are simply multiplied, and show that this mispricing persists even when the underlying legs are themselves well-calibrated — pointing to a markup applied at the parlay-pricing stage rather than to compounded leg-level distortion.

\paragraph{Systematic biases in prediction and sports-betting markets} Against the efficiency baseline, a substantial empirical literature documents systematic deviations. The most extensively studied is the favorite-longshot bias. In many betting markets, contracts on longshots are overpriced relative to their empirical winning frequencies, and contracts on favorites underpriced~\cite{thaler1988parimutuel,ottaviani2008favorite,snowberg2010explaining}. The direction of the bias is not universal — a reverse favorite-longshot bias has been documented in MLB and NHL money-line markets and in some US point-spread markets, with the direction depending on whether the favorite's win probability lies above or below one half~\cite{newall2021favorite,whelan2024risk}. Proposed explanations include risk-loving preferences over longshots~\cite{snowberg2010explaining}, bookmaker behavior in the presence of insider traders~\cite{shin1991optimal}, and bookmaker risk aversion in competitive fixed-odds markets~\cite{whelan2024risk}. Bürgi et al.~\cite{burgi2026makers} document a favorite-longshot bias on Kalshi using 300K transaction-level contracts across all market categories; we differ by stratifying on time-to-expiry within sports, which reveals non-monotonic calibration dynamics that aggregate analyses smooth over. A separate, behaviorally motivated thread relates calibration distortions to non-linear probability weighting of the kind formalized by Tversky and Kahneman~\cite{tversky1992advances} and Prelec~\cite{prelec1998}, the latter of whom derives the compound-invariant family $w(p) = \exp\{-({-\ln p})^\alpha\}$ axiomatically; we adopt this functional form for our near-expiry calibration fits in Section~\ref{sec:tte-miscal}. Across both threads, however, calibration is treated as a static, level-dependent property of a market. To our knowledge, no prior work systematically characterizes how the shape of the calibration function varies with time-to-expiry within event-contingent markets. Filling that gap is the empirical contribution of this paper.

\section{Data and Definitions}\label{sec:data}
This section describes the trade-level data used throughout the paper and formalizes the calibration objects we estimate in Sections~\ref{sec:tte-miscal}.

\subsection{Data}\label{subsec:data}
We obtain trade-level data from Kalshi, the largest CFTC-regulated event-contract exchange in the United States, covering the regular-season moneyline markets of three major North American sports leagues: the National Basketball Association (NBA), Major League Baseball (MLB), and the National Hockey League (NHL). A moneyline market on a given game is a pair of binary contracts that pay \$1 if the designated team wins outright and \$0 otherwise; by construction at each time the two contract prices sum to one, modulo the bid-ask spread.

Our sample spans early March through mid-May 2026. The window is bounded on the right by the end of our data feed and on the left by the start of each league's covered season activity, which yields 11~weeks of trades for the NBA, 10~weeks for the NHL, and 7~weeks for the MLB. After excluding all canceled and voided markets, the sample contains approximately 23~million executed moneyline trades: 13{,}009{,}643 for the NBA, 7{,}148{,}254 for the MLB, and 2{,}819{,}289 for the NHL. Each trade record consists of a contract identifier, an execution timestamp, an executed price (in cents), an executed size, and — once the underlying game settles — a binary outcome indicator. Settlement is determined by the platform; we do not re-derive outcomes from external sources.

We let $i$ index trades, write $p_i \in (0, 1)$ for the executed price expressed as a probability (cents divided by 100), and write $Y_i \in \{0, 1\}$ for the realized outcome of the contract on which trade $i$ was executed. All reported prices are contract execution prices and exclude exchange
fees. Fees are charged separately, on top of the trade price.

\subsection{Time to Expiry}\label{subsec:tte}
For each trade $i$, we define \textbf{time to expiry} (TTE) as the time elapsed between the trade timestamp and the market closing time of the underlying contract:

\begin{align*}
    \tau_i \;=\; t^{\text{close}}_i \;-\; t^{\text{trade}}_i \;\geq\; 0.
\end{align*}

Here, $t^{\text{close}}$ denotes the exchange's \emph{market close time}---the scheduled end of trading, which for our sports markets coincides with the end of the game. We do not use the venue's later administrative settlement timestamp.

We measure $\tau$ in minutes throughout the paper. A trade with $\tau_i$ large corresponds to early speculation well before a game begins; a trade with $\tau_i$ small corresponds to in-game or late-game activity. TTE is the central conditioning variable of the paper. 

We group trades into five TTE buckets with right-open intervals

\begin{align*}
    B_1 = [0, 10),\; B_2 = [10, 30),\; B_3 = [30, 90),\; \\
    B_4 = [90, 240),\; B_5 = [240, \infty),
\end{align*}

all in minutes. The boundaries are approximately log-spaced and were chosen to give finer resolution near expiry, where we anticipate the most rapid evolution of calibration, while still placing a non-trivial number of trades in each bucket. We summarize each bucket by its midpoint (5, 20, 60, 165, and 360~minutes respectively) when plotting parameter trajectories against $\tau$; the rightmost bucket is unbounded above, and we treat its midpoint as 360~minutes for visualization only.

\subsection{Calibration}\label{subsec:calib}
We adopt the standard definition of the calibration function of a probability forecast. For a given (possibly conditional) set of trades, calibration is the conditional expectation of the outcome given the price:
\begin{align*}
    C(p)=\mathbb{E}[Y|P=p]
\end{align*}

A forecast is \textit{perfectly calibrated} if $C(p) = p$ for all $p$ in the support, i.e., contracts trading at a price of $p$ resolve favorably with empirical frequency $p$. Deviations from the diagonal $C(p) = p$ are the object of interest. Empirical estimation of $C$ on a finite sample is performed by grouping trades into price bins and computing the within-bin empirical win rate.

\section{Time-to-Expiry-Conditional Miscalibration}\label{sec:tte-miscal}
This section presents the central empirical result of the paper. For each league we estimate the power-law parameter $\gamma$ and the Platt slope $a$ within each of the five TTE buckets defined in Section~\ref{subsec:tte}, then fit the quadratic-in-log meta-model to characterize how these parameters evolve with time to expiry. We find that both calibration parameters depart sharply from their perfect-calibration reference value of one as expiry approaches, and that this pattern is consistent across all three leagues despite differences in their long-horizon behavior.

\subsection{Estimation}\label{subsec:est}

For each league \(\ell\) and TTE bucket \(b\), we pool all trades across the league's covered weeks and fit two parametric calibration models to the bucketed trades: the power-law \(\hat{C}(p) = p^\gamma / \bigl(p^\gamma + (1-p)^\gamma\bigr)\), and Platt scaling \(\hat{C}(p) = \sigma\bigl(a \operatorname{logit}(p) + b\bigr)\). Both fits are weighted by trade count within price bins. This yields ten parameter estimates per league: five values of \(\hat{\gamma}_{\ell,b}\) and five values of \(\hat{a}_{\ell,b}\), indexed by bucket midpoint \(\tau_b \in \{5,20,60,165,360\}\) minutes.

We then fit the meta-model \(\theta(\tau) = c_0 + c_1 \log \tau + c_2(\log \tau)^2\) separately to each parameter trajectory. Model selection between this quadratic-in-log specification and the restricted linear-in-log alternative \((c_2 = 0)\) is performed by AIC; the quadratic is preferred for both \(\gamma\) and \(a\) across all three leagues, and BIC agrees.

\subsection{Miscalibration Evolving Shape}\label{subsec:miscal-change}

\begin{figure*}[t]
  \centering
  \includegraphics[width=0.48\textwidth]{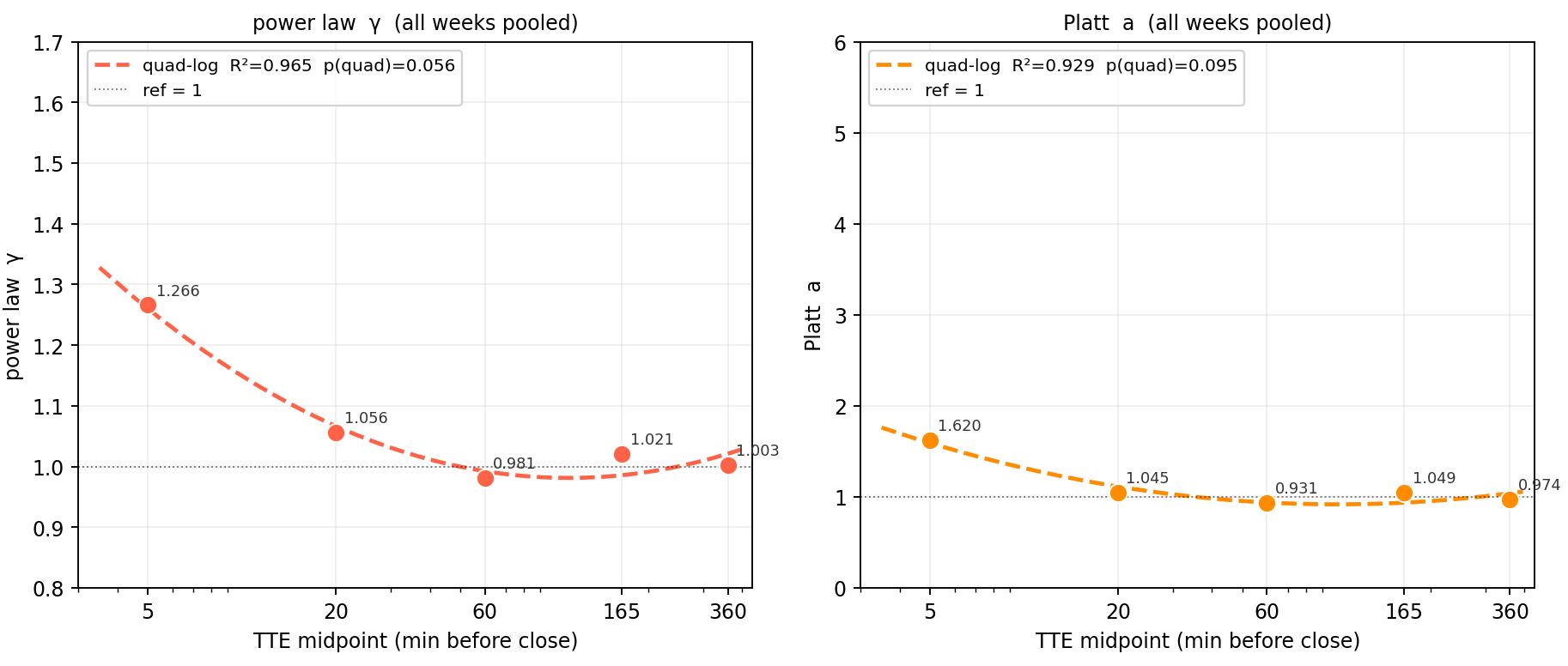}\hfill
  \includegraphics[width=0.48\textwidth]{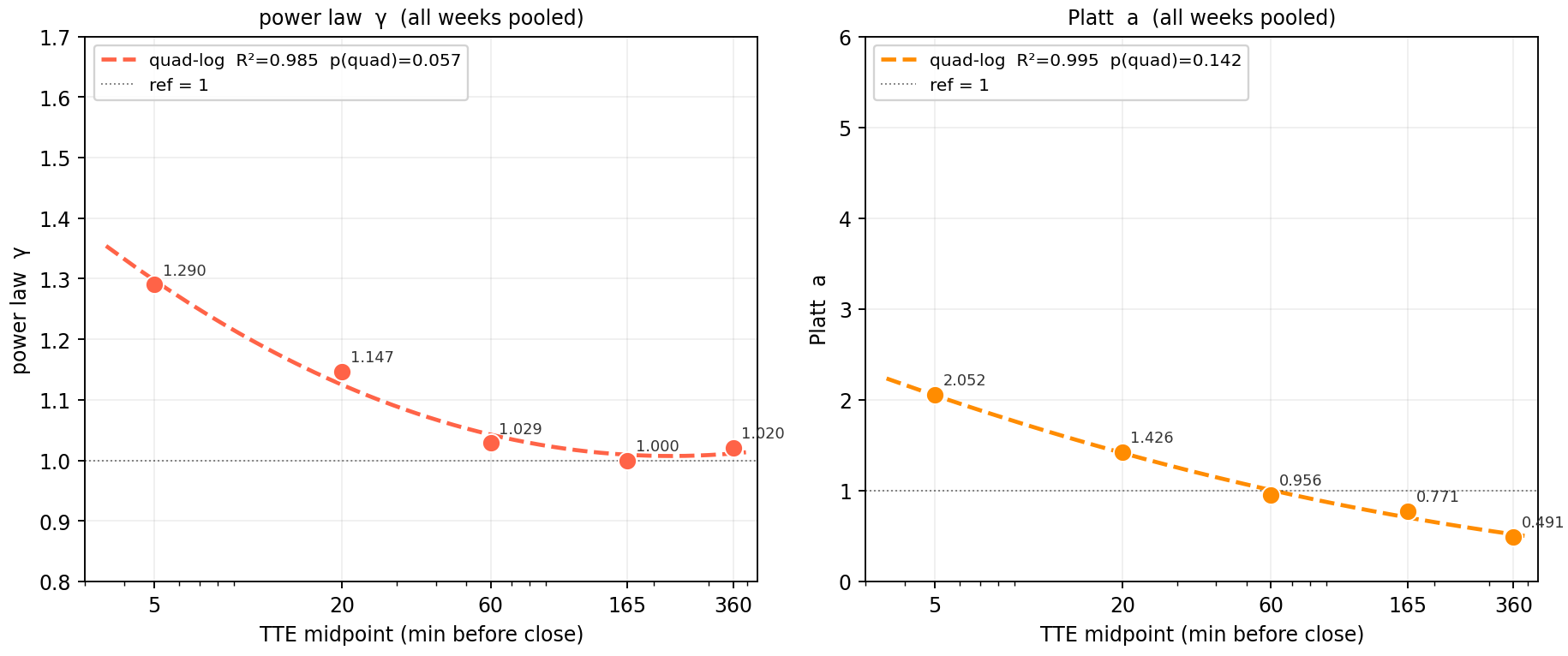}\\[0.5em]
  \includegraphics[width=0.48\textwidth]{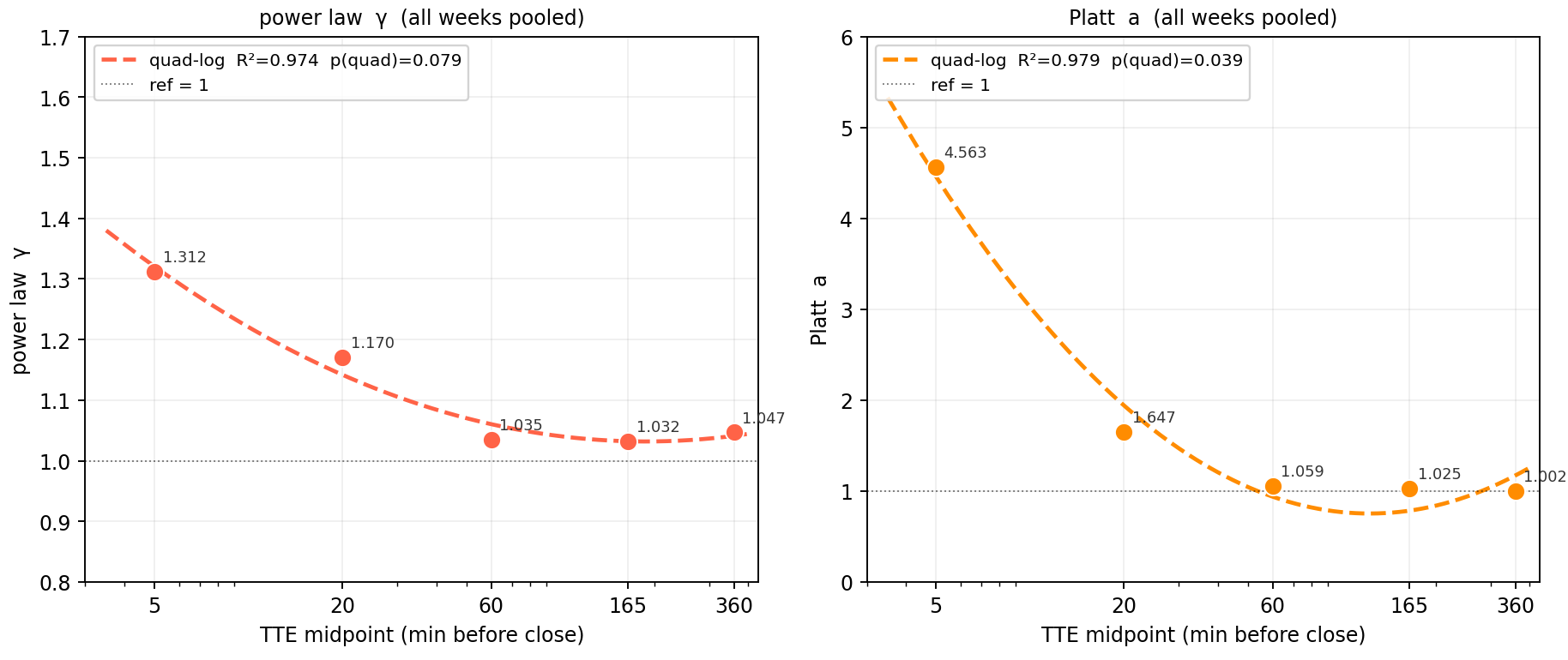}
  \caption{Pooled calibration parameters as a function of time to expiry,
  for the NBA (top left), MLB (top right), and NHL (bottom left). Within
  each panel, the left subplot shows the power-law exponent \(\hat{\gamma}\)
  and the right subplot the Platt slope \(\hat{a}\), each plotted against
  the TTE bucket midpoint in minutes. Dashed reference line at one
  corresponds to perfect calibration. Dashed curve is the fitted
  quadratic-in-log meta-model; legend reports its \(R^2\) and \(p\)-value
  on the quadratic term.}
  \label{fig:tte-trajectories}
\end{figure*}

Figure~\ref{fig:tte-trajectories} reports the parameter trajectories for the NBA, MLB, and NHL together with their fitted quadratic-in-log meta-curves. Two patterns are visible across all three leagues.

\paragraph{First, calibration deteriorates sharply near expiry.} Both \(\hat{\gamma}\) and \(\hat{a}\) are well above their perfect-calibration reference value of one in the smallest TTE bucket. The 5-minute bucket yields \(\hat{\gamma} \in [1.27,1.31]\) across leagues and \(\hat{a}\) ranging from \(1.62\) for the NBA to \(4.56\) for the NHL, with MLB at \(2.05\). Values of \(\gamma\) and \(a\) above one indicate that the calibration function is S-shaped relative to the diagonal: contracts trading at prices near \(0\) resolve favorably \emph{less often} than those prices suggest, and contracts trading at prices near \(1\) resolve favorably \emph{more often}. The magnitude of this distortion in the smallest bucket is substantial---at \(\hat{a} = 4.57\) for NHL, a contract trading at \(p = 0.4\) has an empirical win rate near zero in our data, far below any plausible probability interpretation of the price.

\paragraph{Second, both calibration models agree on the qualitative pattern.} Across all three leagues, \(\hat{\gamma}\) and \(\hat{a}\) co-vary: when one is far from its reference value, so is the other. This is reassuring because the two models impose different functional forms on the calibration curve, and a distortion visible in only one would be suspect as a model artifact rather than a property of the market.

The meta-fits are tight: quadratic-in-log \(R^2\) values exceed \(0.92\) across all six trajectories, and AIC and BIC consistently prefer the quadratic over a linear-in-log alternative.

\subsection{Inside the Calibration Curve}\label{subsec:calib-curve}

\begin{figure*}[t]
  \centering
  \includegraphics[width=\textwidth]{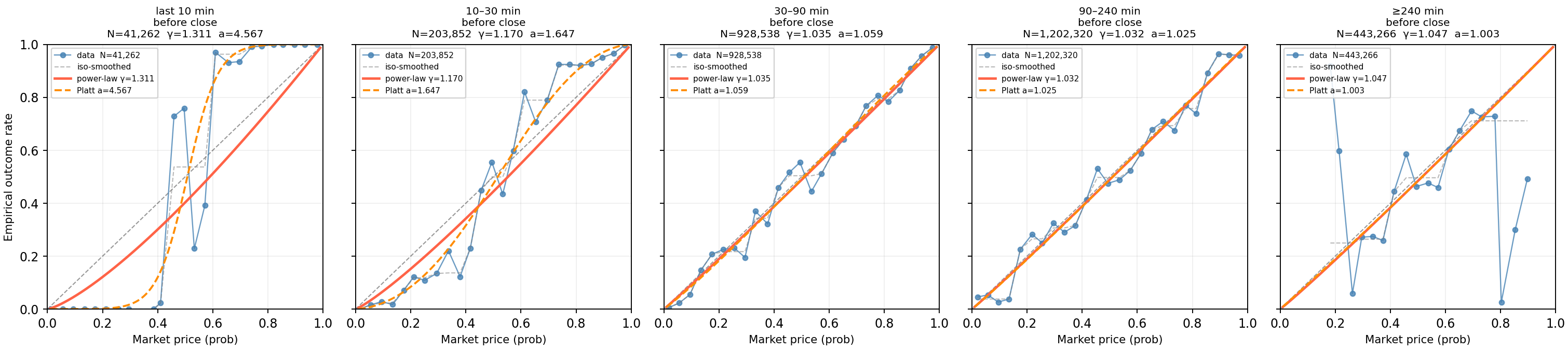}
  \caption{Pooled NHL calibration curves by TTE bucket. Each panel shows
  the empirical outcome rate (blue dots) and isotonic-smoothed curve
  (dashed) as a function of executed market price, with the fitted
  power-law (red) and Platt (orange) overlays. Bucket-level trade
  counts and fitted parameters \(\hat{\gamma}\) and \(\hat{a}\) are
  reported in each panel title. The diagonal corresponds to perfect
  calibration.}
  \label{fig:nhl-buckets}
\end{figure*}

To clarify what the parameter trajectories of Figure~\ref{fig:tte-trajectories} are summarizing, Figure~\ref{fig:nhl-buckets} shows the underlying empirical calibration curves for the NHL in each of the five TTE buckets, with the fitted power-law and Platt curves overlaid.

In the 30--90 and 90--240 minute buckets, the empirical calibration curve is visually indistinguishable from the diagonal across the entire \((0,1)\) price range. In the middle of the TTE range, the market is well-calibrated.

In the 0--10 minute bucket, the calibration curve departs from the diagonal not as a smooth S-shape but as something closer to a step function. The empirical outcome rate sits at essentially zero across prices below \(0.4\)---contracts trading at \(10\)--\(40\) cents in the final ten minutes resolve favorably almost never---transitions sharply around \(p \approx 0.5\), and saturates near one across the upper half. The shape more closely resembles a Prelec-style probability-weighting distortion with high curvature than the smooth S-shapes the power-law and Platt families can produce. Across all three leagues the asymmetry is consistent: large distortion approaching expiry, small or noisy distortion long before it. This near-expiry concentration is the empirical anchor for the behavioral interpretation we develop next.

\begin{figure*}[t]
  \centering
  \includegraphics[width=\textwidth]{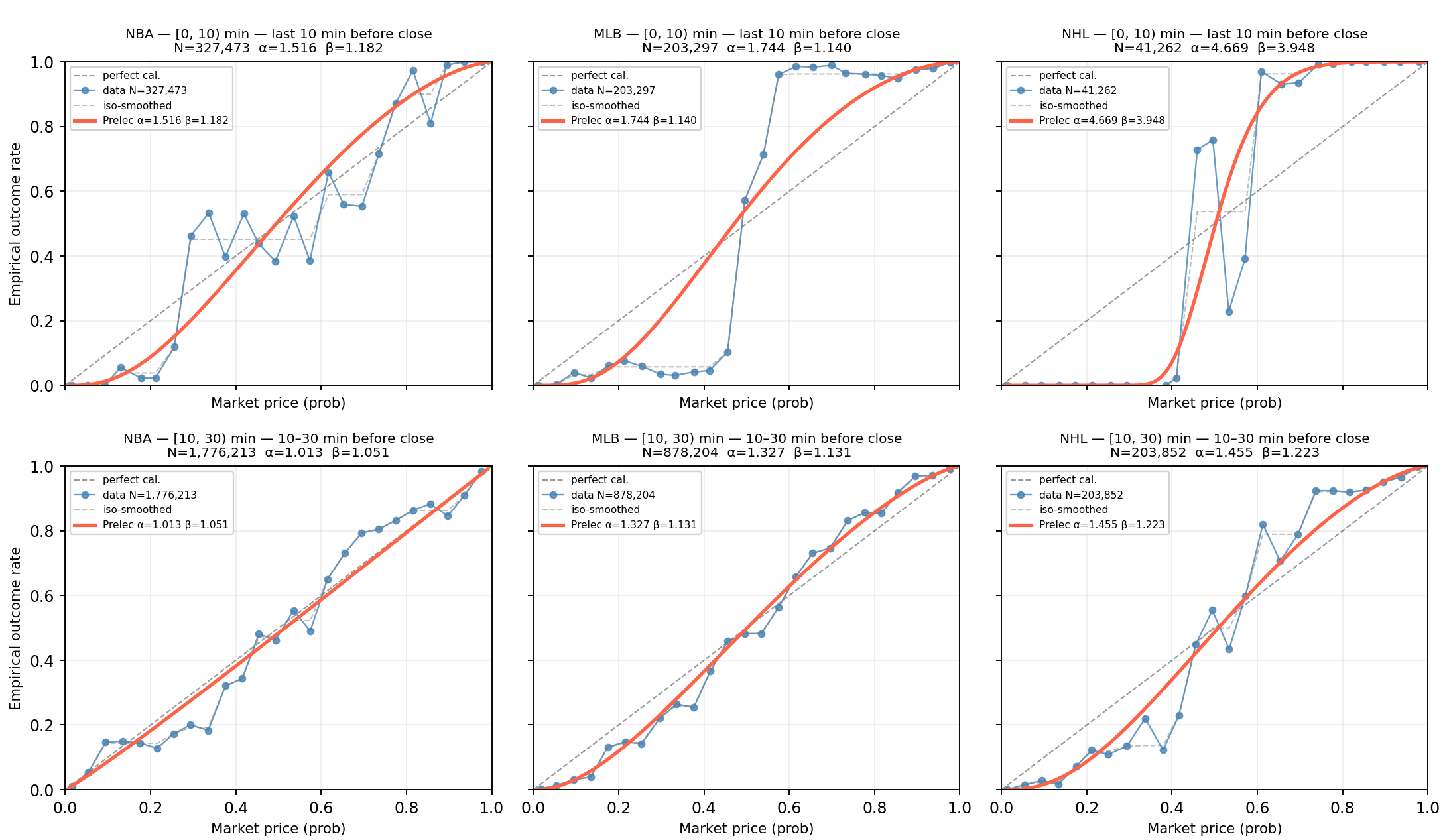}
  \caption{Prelec-II fits in the two TTE buckets nearest expiry, for all
  three leagues. Top row: 0--10 minute bucket. Bottom row: 10--30 minute
  bucket. Each panel shows the isotonic-smoothed empirical calibration
  curve (dashed) and the fitted Prelec-II curve (red), with fitted
  \(\hat{\alpha}\) and \(\hat{\beta}\) in the panel title. The diagonal
  corresponds to perfect calibration.}
  \label{fig:prelec-closing}
\end{figure*}

\subsection{The Near-Expiry Distortion as Probability Weighting}\label{subsec:tte-prelec}

The 0--10 minute calibration curves in Figure~\ref{fig:nhl-buckets} exhibit a sharpness at the midpoint that the power-law and Platt families cannot reproduce. We fit the two-parameter Prelec-II weighting function \cite{prelec1998}, \(w(p) = \exp\{-\beta(-\ln p)^\alpha\}\), with \(\alpha,\beta > 0\), to the isotonic-smoothed empirical calibration curve in each \((\mathrm{league}, \mathrm{bucket})\) cell. Perfect calibration corresponds to \((\alpha,\beta) = (1,1)\); the fixed point \(p^\star\) is determined jointly by both parameters and is not pinned to the canonical \(1/e\) of the one-parameter Prelec-I form.

Figure~\ref{fig:prelec-closing} reports the fits for the two TTE buckets nearest expiry. The Prelec distortion is not a constant feature of these markets but emerges sharply as expiry approaches. Section~\ref{subsec:miscal-change} established that across the bulk of the TTE range — from hours before settlement down to roughly half an hour — calibration parameters sit near their perfect-calibration reference values, and the empirical curve coincides with the diagonal. By the 10–30 minute bucket a mild Prelec shape has begun to form. In the final 10 minutes, the Prelec fit tracks a step-like curve. Whatever mechanism produces the distortion therefore activates only in the closing minutes of trading — a demand component absent for the bulk of a contract's life and dominating its final moments.

The direction of the fitted distortion is the inverse of the canonical probability-weighting result. Lottery-choice settings typically yield \(\hat{\alpha} < 1\) \cite{tversky1992advances, prelec1998}. Our near-expiry fits yield \(\hat{\alpha} > 1\) and the opposite S-shape: prices in the interior of \((0,1)\) are pulled away from \(1/2\) toward \(0\) or \(1\). This is not the prospect-theory picture of a gambler overweighting a longshot; it is the picture of a risk-averse trader treating a moderately unfavorable position as nearly certain to lose, and paying a premium on the opposing side to escape it. The magnitude of the distortion varies across leagues in the order NHL \(>\) MLB \(>\) NBA, consistent with the degree to which late-game outcomes resolve in discrete versus continuous fashion.

We interpret the near-expiry Prelec shape as a market-level signature of insurance demand: traders holding losing positions in the final minutes face convex losses that make paying above the fair price on the opposing side rational under expected utility, even though it is irrational under probability matching. Aggregated, this hedging activity compresses the calibration curve toward the step shape we observe. Alternative mechanisms---liquidity withdrawal by informed participants, entry of uninformed late traders, herding on visible game state---could produce a qualitatively similar pattern, and distinguishing them requires order-book and trader-identity data beyond our scope. What our fits establish is the shape and magnitude of the distortion itself: in the final ten minutes before settlement, sports prediction-market prices deviate from probabilities by an amount that is large, systematic, and consistent across three major leagues.

\section{Parlay Pricing: Markup in a Calibrated Regime}\label{sec:parlay-miscal}
The first half of this paper established that single-leg moneyline prices on Kalshi are well-calibrated in the middle of the TTE range and exhibit a sharp Prelec-like distortion only in the final minutes before settlement. We now turn to the prediction market's natural derivative product: the parlay. A parlay is a single contract whose payoff depends on the joint occurrence of $n$ underlying events. Under independence, the fair price of an $n$-leg parlay is the product of its leg prices. Treating prediction-market prices as probabilities, a downstream user — or a market-maker — who priced a parlay this way would be acting consistently with the foundational claim under which the single-leg market operates.

We show that this consistency does not hold empirically. Cross-game parlays on Kalshi are systematically priced above the product of their contemporaneous leg prices, and the magnitude of the overpricing grows with leg count. The legs themselves are drawn from the TTE range in which Section~\ref{sec:tte-miscal} established that single-leg calibration is essentially perfect, so the overpricing cannot be attributed to leg-level distortion compounding through the product. The deviation originates at the parlay-pricing stage.

\subsection{Data}

The parlay sample is drawn from Kalshi's parlay trades over a 15-day window, April 29 through May 13, 2026, sitting within the moneyline window of Section~\ref{subsec:data}. For each parlay execution at time \(t_i\), we match each constituent leg to the most recent execution price in the corresponding single-game contract at or before \(t_i\). We retain a parlay only if every leg has an eligible execution within \(300s\) of the parlay timestamp. Starting from a base of \(153{,}173\) parlay trades for which a contemporaneous leg-price independence product can be computed, we apply three filters designed to isolate parlays whose underlying legs lie in the calibrated regime identified in Section~\ref{sec:tte-miscal}. (i) Every leg of the parlay must have time-to-expiry between \(30\) and \(240\) minutes at the time of the parlay trade. By Section~\ref{sec:tte-miscal}, this is the TTE range over which single-leg calibration parameters sit at their perfect-calibration reference values across all three leagues; restricting to this range means leg-level miscalibration cannot contribute to any overpricing we observe. (ii) We restrict to cross-game parlays, i.e., parlays whose legs reference outcomes of distinct games, excluding same-game combinations where leg outcomes are non-trivially correlated. After these filters the analysis sample contains \(N = 12{,}639\) parlay trades spanning two to eleven legs.

For each parlay trade \(i\) with \(n_i\) legs, we observe the executed parlay price \(P^{\mathrm{exec}}_i \in (0,1)\) and the contemporaneous executed prices \(\{p_{i,1},\dots,p_{i,n_i}\}\) of its component legs. The \textbf{independence-product} price is \(P^{\mathrm{ind}}_i = \prod_{j=1}^{n_i} p_{i,j}\), the fair price of the parlay under the joint hypothesis that (a) leg prices are probabilities and (b) legs are independent. The \textbf{overpricing ratio} is \(R_i = P^{\mathrm{exec}}_i / P^{\mathrm{ind}}_i\). A value of \(R = 1\) corresponds to parlay pricing consistent with the price-as-probability interpretation under independence; \(R > 1\) corresponds to overpricing.

\subsection{Overpricing Grows with Leg Count}\label{subsec:parlay-overprice}

\begin{figure*}[t]
  \centering
  \includegraphics[width=0.8\textwidth]{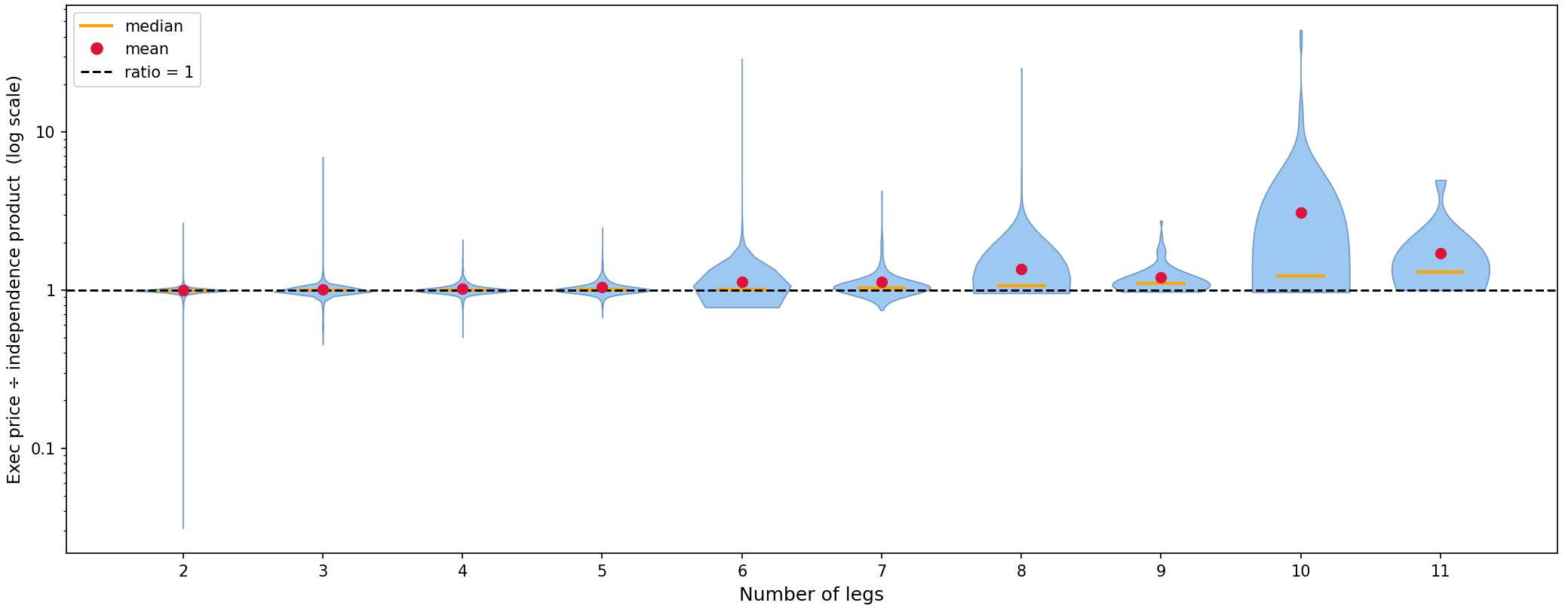}
  \caption{Parlay overpricing ratio
  \(R = P^{\mathrm{exec}} / P^{\mathrm{ind}}\) by leg count, on a
  log-vertical axis. Each violin shows the distribution of \(R\) for
  parlays with a given number of legs in the analysis sample. Orange
  bar: median. Red dot: mean. Dashed reference line at \(R = 1\)
  corresponds to parlay pricing consistent with the price-as-probability
  interpretation under leg independence.}
  \label{fig:parlay-overpricing}
\end{figure*}

Figure~\ref{fig:parlay-overpricing} reports the overpricing ratio \(R\) as a function of leg count, on a log-vertical axis. Table~\ref{tab:parlay_overpricing} reports the underlying statistics. We focus on the median, both because the distribution of \(R\) is heavy-tailed at large \(n\) and because the median is the directly interpretable summary of ``a typical parlay's overpricing factor.''

\begin{table}[h]
\centering
\caption{Overpricing ratio \(R = P^{\mathrm{exec}} / P^{\mathrm{ind}}\) by leg count, on cross-game parlays whose every leg has TTE between 30 and 240 minutes.}
\label{tab:parlay_overpricing}
\begin{tabular}{rrrrrr}
\hline
Legs & \(N\) & Mean & Median & Std & 95\% CI (mean) \\
\hline
 2 & 5{,}846 & 0.994 & 0.991 & 0.071 & [0.992, 0.996] \\
 3 & 3{,}111 & 1.006 & 0.995 & 0.173 & [1.000, 1.012] \\
 4 & 1{,}523 & 1.022 & 0.999 & 0.108 & [1.016, 1.027] \\
 5 &   928 & 1.040 & 1.005 & 0.134 & [1.031, 1.049] \\
 6 &   584 & 1.120 & 1.013 & 1.176 & [1.024, 1.215] \\
 7 &   371 & 1.122 & 1.037 & 0.299 & [1.092, 1.153] \\
 8 &   140 & 1.353 & 1.066 & 2.063 & [1.010, 1.696] \\
 9 &    67 & 1.199 & 1.105 & 0.305 & [1.126, 1.273] \\
10 &    57 & 3.078 & 1.223 & 7.439 & [1.130, 5.026] \\
11 &    12 & 1.712 & 1.305 & 1.050 & [1.092, 2.332] \\
\hline
All & 12{,}639 & 1.028 & 0.996 & 0.632 & [1.017, 1.039] \\
\hline
\end{tabular}
\end{table}

Three patterns are visible. First, at low leg counts \((n \in {2,3,4})\), the median overpricing ratio sits essentially at one---\(0.991\), \(0.995\), \(0.999\)---and parlay prices are consistent with the independence-product. Second, beyond approximately \(n = 5\), a clear and monotonic growth in median overpricing begins; \(1.01\) at five legs, \(1.07\) at eight legs, \(1.22\) at ten legs, and \(1.31\) at eleven legs. Third, dispersion grows substantially at high leg count, with the means at \(n = 8\) and \(n = 10\) reflecting heavy upper tails. We treat the \(n > 10\) cells as small-sample evidence of the same trend rather than as separately calibrated estimates. Count \(N\) falls below \(70\) in these cells and the heavy right tails make the means unstable. The median, more robust to tails, continues to grow smoothly across the range.

A simple regression captures the rate. Fitting \(\log \tilde{R}(n) = \beta_0 + \beta_1 n\) to the median overpricing ratios across leg counts \(n = 2\) through \(11\) yields \(\hat{\beta}_1\approx 0.029\) \((R^2 \approx 0.94)\) corresponding to a per-leg multiplicative inflation of roughly \(3\%\) in the median. The much larger \(\hat{\delta}\) values in the headline figure (Figure~\ref{fig:headline}) come from a parametric model fit to trade-level outcomes rather than to the median ratios; the two parameterizations summarize the same phenomenon under different functional forms.

\subsection{Calibration of the Parlay Market}\label{subsec:parlay-market-calib}

\begin{figure*}[t]
  \centering
  \includegraphics[width=\textwidth]{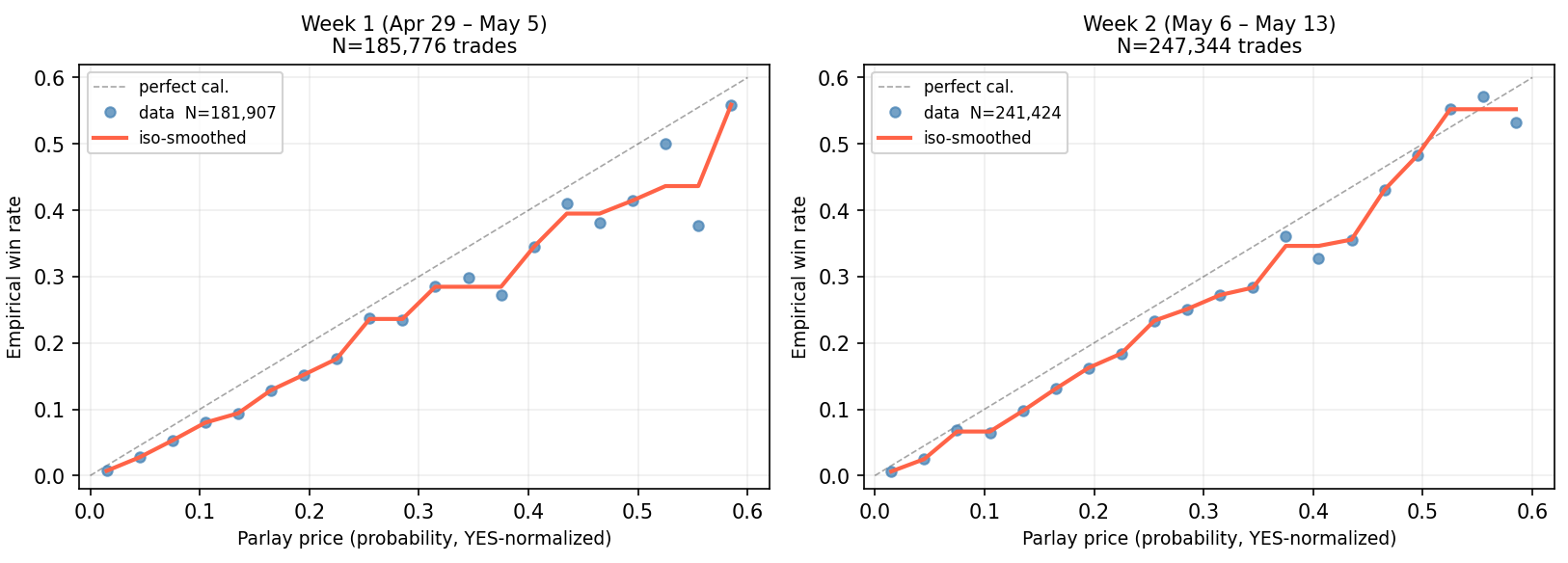}
  \caption{Empirical calibration of cross-game parlay trades, pooled by week of the analysis window. Left: Week 1 (Apr 29 -- May 5). Right: Week 2 (May 6 -- May 13). Both YES-side and NO-side trades are included; NO-side prices are reflected to the YES perspective for pooling. In each panel the empirical win rate (blue dots) and isotonic-smoothed curve (red) are plotted against the executed parlay price. Axes are confined to prices below \(0.6\), containing \(98\%\) of trades in the sample. The diagonal corresponds to perfect calibration.}
  \label{fig:parlay-calibration}
\end{figure*}

The overpricing-vs-leg-count result of Section~\ref{subsec:parlay-overprice} captures one dimension of the parlay markup; the calibration curve of the parlay product captures another. Figure~\ref{fig:parlay-overpricing} reports the empirical calibration curve of cross-game parlay trades, pooled weekly across the analysis window. Following the convention of Section~\ref{sec:tte-miscal}, we plot the empirical win rate as a function of executed parlay price. We confine the axis to prices below \(0.6\), which contains \(98\%\) of trades in the sample.

The pattern is qualitatively distinct from the single-leg curves of Section~\ref{sec:tte-miscal}. Both weekly curves sit consistently \emph{below} the diagonal across essentially the entire observed price range, with the empirical win rate running between two and ten percentage points below the price. A parlay priced at \(0.30\) wins on average about \(24\%\) of the time; one priced at \(0.40\), about \(30\%\). The shape of the deviation is a roughly uniform downward shift---consistent with a flat or near-flat additive markup over fair pricing rather than a price-dependent behavioral distortion.

\subsection{Discussion}

Three observations narrow the possible explanations for the overpricing. (i) Leg-level calibration is essentially perfect in the TTE regime from which parlay legs are drawn, as shown in Section~\ref{sec:tte-miscal}, so leg-level distortion cannot account for the parlay overpricing. (ii) The cross-game restriction means leg outcomes can reasonably be treated as independent, so the overpricing cannot be attributed to a correlation premium of the kind that explains the well-documented house-edge inflation in same-game parlays \cite{wizardOfOddsSGP}. (iii) The calibration curve at the parlay level is shifted uniformly below the diagonal, as shown in Section~\ref{subsec:parlay-market-calib}, inconsistent with a price-dependent behavioral weighting and consistent with an additive markup.

Markups are well-documented in the conventional sportsbook setting, where house edges on multi-leg products are observed to compound substantially above per-leg vig \cite{hegarty2025margins}, and have recently been documented on Kalshi specifically for low-priced single contracts \cite{burgi2026makers}. Our results extend this picture to the parlay product: even when individual leg prices are unbiased estimators of outcome frequencies, the parlay built from those legs carries a parlay-level pricing wedge whose effective magnitude grows multiplicatively with leg count.

Whether this markup is best modeled as a fee structure, a capital-reserve premium against joint-probability estimation risk \cite{rana2026parlaymarket}, or demand-driven overpayment by traders seeking lottery-like payoffs \cite{whelan2026parlaypuzzle} cannot be distinguished by our data alone. What our results establish is: cross-game parlay prices on Kalshi are not the prices the standard price-as-probability interpretation implies. The deviation is systematic, grows with leg count, takes the form of a roughly uniform downward shift in the calibration curve, and exists in a regime where the underlying legs are calibrated.

\section{Conclusion}\label{sec:conclusion}

Prediction market prices are increasingly treated as probabilities---by traders, researchers, journalists, and the designers of derivative products built atop these markets. This paper has examined that interpretation along a dimension that has received little prior attention, time to expiry, and has documented two distinct ways in which it fails systematically.

First, within a single contract's life, calibration is not static. Fitting power-law and Platt-scaling calibration models within TTE buckets for moneyline trades on the NBA, MLB, and NHL, we find that both parameter trajectories depart sharply from their perfect-calibration reference values as expiry approaches, while sitting at those values in the middle of the TTE range. In the final ten minutes before settlement the empirical calibration curve becomes step-like, fitting a Prelec form with curvature parameter \(\hat{\alpha}\) well above one---the opposite sign of the canonical lottery-choice fit, consistent with insurance-demand behavior by traders holding losing positions.

Second, this leg-level calibration in the middle of the TTE range does not transfer to derivative products. Cross-game parlays on Kalshi, drawn from a sample whose legs fall entirely within the calibrated regime, are systematically overpriced relative to the product of their contemporaneous leg prices. Because the underlying legs are calibrated and the cross-game restriction precludes a correlation premium, the deviation originates at the parlay-pricing stage.

Together, these findings establish that the price-as-probability interpretation fails in qualitatively different ways at different points in a contract's life and at different levels of product complexity. Both deviations are systematic: TTE-conditional in the first case, leg-count-dependent in the second. Both admit computational correction and quantifying the out-of-sample value of such corrections is the natural next step. Any application of prediction-market prices as probability inputs requires conditioning on both time to expiry and product type, not on price alone.

\end{document}